# The Potsdam MRS Spectrograph – heritage of MUSE and the impact of cross-innovation in the process of technology transfer

B. Moralejo*[a], M.M. Roth[a], P. Godefroy[b], T. Fechner[a], S.M. Bauer[a], E. Schmälzlin[a], A. Kelz[a], R. Haynes[a]

[a]Leibniz-Institut für Astrophysik Potsdam, innoFSPEC, An der Sternwarte 16, 14482 Potsdam, Germany; [b]Winlight System S.A., rue Benjamin Franklin, ZA St Martin, 84120 Pertuis, France

## ABSTRACT

After having demonstrated that an IFU, attached to a microscope rather than to a telescope, is capable of differentiating complex organic tissue with spatially resolved Raman spectroscopy, we have launched a clinical validation program that utilizes a novel optimized fiber-coupled multi-channel spectrograph whose layout is based on the modular MUSE spectrograph concept. The new design features a telecentric input and has an extended blue performance, but otherwise maintains the properties of high throughput and excellent image quality over an octave of wavelength coverage with modest spectral resolution. We present the opto-mechanical layout and details of its optical performance.

## 1. INTRODUCTION

After first tentative experiments to utilize astronomical instrumentation for the purpose of minimal invasive optical diagnostics in medicine[1], we have been able to demonstrate that the technique of integral field spectroscopy (IFS) from astronomy, applied to a microscope rather than to a telescope, is indeed capable of differentiating complex organic tissue on the basis of spatially resolved Raman spectroscopy[2]. Raman light scattering has emerged as a powerful tool for chemical characterization owing to recent improvements in laser and light detection technologies. Raman spectroscopy essentially means to illuminate an area of a sample with monochromatic light and then to detect the shift in energies of the inelastically scattered light. As these shifts depend on the rotational and vibrational degrees of freedom of the molecules that constitute the sample, this technique provides a minimal invasive diagnostic, e.g. for organic tissue. Like in astronomy, when spectra at different positions of an object are collected, spectral maps (chemical maps in case of Raman spectroscopy), can be created. In medicine and life science, this is usually a time-consuming stepwise process since each spectrum is recorded individually and a fraction of the time is invested in mechanical displacements of the measuring head. This is a particular disadvantage when working with organic samples in Raman spectroscopy because long laser exposures can damage the sample and photo-bleaching could induce alterations in organic properties.

The IFS novel approach has attracted the interest of the bio-photonic community on the grounds that IFS is orders of magnitude more efficient than current state-of-the-art Raman instrumentation. It has a promising potential to revolutionize resection margin identification in cancer diagnostics[3] and surgery. Based on these findings, we have launched a joint research program with clinical experts and industrial partners where astrophysicists and life science specialists work closely to develop an optimized fiber-coupled multi-channel spectrograph with a dedicated novel integral field unit (IFU) for the purpose of a formal validation program in dermatology towards a future prototype for clinical applications. On the basis of the concept of modular and replicable spectrographs for MUSE[4], we have designed and built a modified spectrograph that, as opposed to its predecessor, is adapted to a telecentric fiber input and has an extended blue performance, but otherwise maintains the properties of high-throughput and excellent image quality over an octave of wavelength coverage with modest spectral resolution. Owing to the immediate purpose of the experiment, the spectrograph is dubbed the Multiplex-Raman-Spectrograph (MRS). However, given the performance of the new system, and building on the successful concepts of replicable spectrographs and small series production, e.g. the MUSE and VIRUS instruments, we also envision to deploy copies for integral field and/or multi-object spectroscopy at astronomical telescopes. We stress the benefits of a partnership between astronomy, other disciplines in academia, and industry as a win-win situation for the expensive development of instrumentation.

*bmoralejo@aip.de

## 2. INSTRUMENT OVERVIEW

The new Potsdam MRS is a fully refractive fiber-fed optical system derived from the MUSE spectrograph concept. Initially developed for spatially resolved Raman spectroscopy, is also suitable for IFS and multi-object spectroscopy (MOS) in astronomy. With the aim to be adapted for Raman imaging, some modifications were applied to both optical and mechanical layouts. The slicing mirrors and fore-optics subsystems used for MUSE were replaced by a straight fiber optic array structure of up to 118 mm length (+59 mm to -59 mm with respect to the center of the optical axis marked as 0 mm) that is coupled to a first collimator lens. The input f-number is f/4.33, and the beam at the pupil position exhibits now a non-anamorphic circular aspect rather than elliptical. The wavelength coverage was expanded to the blue part of the spectrum down to 0.35 μm while the red part ends at 0.9 μm. At the detector side, a 4k × 4k CCD manufactured by e2V enhanced with a graded index anti-reflection coating is recording spectra. The linear dispersion is 0.134 nm per pixel, resulting in an estimated resolving power of about 1200…3000 (blue/red) for a 50 μm fiber core.

**Optical design**

The general aim for the Potsdam MRS is to modify the MUSE opto-mechanical layout to accommodate an optical fiber feed, keeping properties of high-throughput and image quality, and improving performance at blue wavelengths, while minimizing changes in the spectrograph itself. In this design, a straight array of telecentric optical fibers is attached to a first silica plane-concave lens by means of an index matching gel at the collimator side of the instrument. The maximum allowed number of input fibers is determined by the length of the input slit and the output diameter of the optical fibers, e.g. about a thousand silica multimode fibers model FVP050055065 (Polymicro Technologies LLC, Phoenix, USA) . In the current setup, a total of 421 identical fibers (114 μm core / 125 μm cladding / 153 μm buffer) compose the pseudo-slit; 400 fibers belong to the squared shape fiber bundle IFU (20 × 20 fibers), and the remaining 21 fibers are available for custom purposes, e.g. testing frequency combs and custom devices. The ZEMAX optical layout is illustrated in Figure 1. Ray tracing calculates the beam path along the spectrograph from an input object until its spectra impinge at the image plane for different objects and wavelengths.

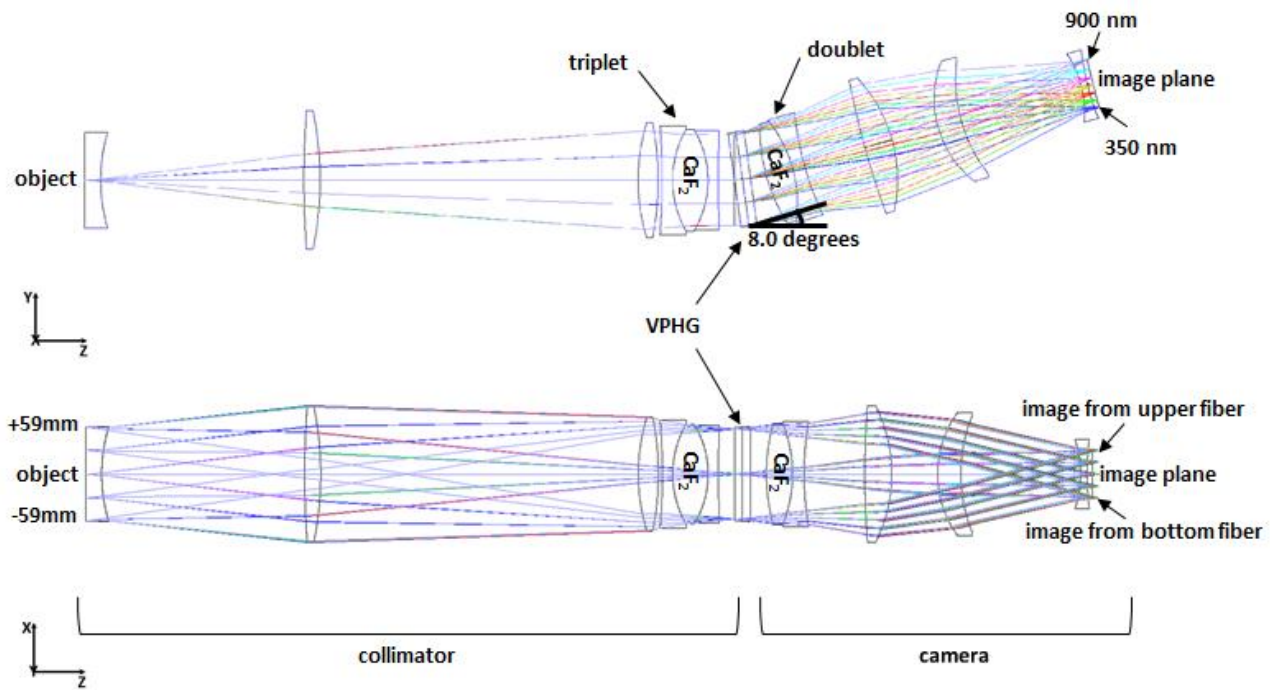

Figure 1. Side-view (top) and top-view (down) of the Potsdam MRS optical layout (spatial and spectral dimensions are specified for X and Y direction, respectively).

The collimator subassembly is formed by three single lenses and one triplet placed just before the grating. Between the collimator and camera subassemblies, a volume phase holographic grating (VPHG) diffracts the collimated input light.

Diffracted transmitted light passes through the camera subassembly and is focused on the image plane where the CCD detector is located. At the other side, two single lenses, a doublet immediately after the VPHG and an aspheric field lens made of silica form the camera part. Both, the triplet at the collimator and the doublet at the camera side, were cemented using RTV141 to avoid stress within the glasses as well as light absorption within the spectral range. To improve performance and increase throughput at the blue part of the spectrum, cemented lenses include a calcium fluoride ($CaF_2$) element. After light passes through the whole system, the resulting spectra of all input objects created within the input field are imaged in a focus plane located perpendicularly to the optical axis. As a feature inherited from MUSE, the field lens at the end of the spectrograph is assembled together with the camera housing providing of vacuum isolation to the dewar and at the same time providing the best image quality at a precise focus distance.

Figure 2 depicts the theoretical averaged transmittance of glasses through all lenses along the spectrograph. Values were calculated with Zemax from the AR coatings and internal reflectance of glasses, however the VPHG diffraction performance is not considered. Transmission in the blue has been enhanced due to the use of $CaF_2$ glasses and predicted values are between 73 – 93 % at the range of 0.35 µm and 0.4 µm while keeping transmission above 93 % for longer wavelengths.

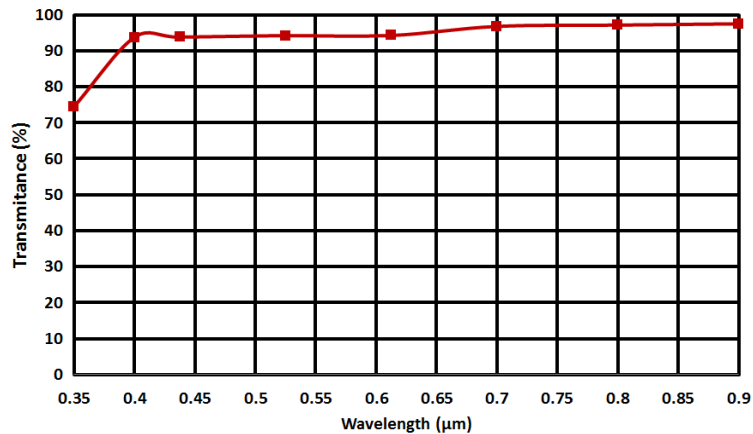

Figure 2. Theoretical transmittance of the Potsdam MRS spectrograph calculated with ZEMAX without considering VPHG characteristic.

A study of unwanted diffused light carried out with Zemax showed that ghost images from any fiber stay below a ratio per pixel of $10^{-6}$. This theoretical analysis considers all reflections between two diopters as well as at the detector and lens faces. There is an exception when considering ghost resulting from VPHG diffraction orders different than first order. This is the case of blue wavelengths coming from second order which are transferred to the red part of the first order diffraction. This undesirable light cannot be avoided by means of baffling. Therefore, it must be removed by filtering the unwanted part of the input light spectra.

## 2.1 Volume phase holographic grating

The VPHG presents a completely new design due mainly to the remapped covered wavelengths. The current shape of the diffraction gelatin is 118 mm diameter circular while the entire VPHG appears with a square shape of 122 mm side length. The thickness dimension is mostly determined by two BK7 plates of 10 mm each protecting the gelatin. The number of lines at the gelatin was initially calculated to span the wavelength range [0.35 – 0.9] µm at the image plane and resulting in a frequency of 429 lines per mm. The Bragg condition is met when the angle of incidence (AOI) occurs at eight degrees. As part of the optical characterization, the diffracted wavefront measurements reported a maximum value of 0.798 peak-to-valley including all external and internal surfaces and internal irregularities when tested at a wavelength of 0.63 µm.

The diffraction performance was tested at AIP and compared with results provided by the manufacturer KOSI (Kaiser Optical Systems, Inc. Michigan, USA). Several band-pass filters in combination with a halogen lamp resulted in six different wavelengths (0.405 µm, 0.45 µm, 0.508 µm, 0.58 µm, 0.69 µm and 0.83 µm) which were used to quantify first order diffraction efficiency at 8 degrees AOI. The measurements were performed at the center of the grating. Figure 3 shows the comparison between efficiencies specified by the manufacturer and values measured at AIP. The overall shape of the resulting curve is similar, however the curve determined at AIP appears slightly above the manufacturer's data. Higher efficiencies of up to a 5 % were measured at AIP, especially in the red and infrared parts of the spectrum. Nevertheless, efficiency measured in the blue at 0.405 µm and 0.45 µm agreed with the manufacturer and differences of less than 2 % were found. In both cases, the VPHG yields best efficiency diffraction at around 0.55 µm. Moving away from this wavelength, efficiency drops to around 50 % at 0.4 µm at blue and 0.77-0.83 µm at red wavelengths depending on the curve under consideration. Due to the noise of the signal between 0.35 µm and 0.4 µm, the efficiency at this range cannot be accurately estimated.

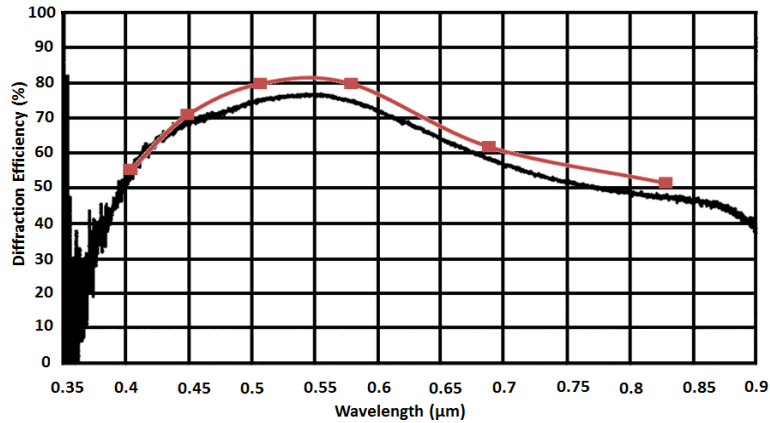

Figure 3. Experimental values of first order diffraction efficiency measured at the center of the VPHG. Black curve represents values measured by KOSI and red curve the values measured at the AIP.

In order to test the spatial homogeneity of the grating as formerly done for MUSE[5], the efficiency at nine different positions along the surface of the grating were compared. This results in efficiency deviations of less than 10% when compared with efficiency at the center. There is an exception at a corner of the grating where efficiency drops by 20% at 0.58 µm wavelength. For the rest of wavelengths at this position, efficiency also drops, however to a lesser extent. When compared to MUSE, these inhomogeneities are less critical. This is because the light emerging from fibers at the VPHG position is defined by a Gaussian shape profile showing a maximum at the center of the grating. Therefore most of the energy is diffracted at the center where efficiency shows the expected behavior.

## 2.2 Mechanical design

The mechanical structure of the spectrograph has been adapted to the updated optical requirements. From the mechanical point of view, four main interfaces can be set at the instrument. The fiber bundle forming a straight line at the input and coupled to the first lens of the collimator defines the slit interface. A second interface is determined by the field lens centered in a barrel and mounted on the camera flange ensuring a fixed interface with the detector assembly. The VPHG interface describes a circular pupil with its external faces parallel to the diffraction grating splitting the spectrograph in collimator and camera subassemblies. Finally, the outer structure that maintains the spectrograph at the right position for storage and ready for operation represents the interface with the entire spectrograph.

The fiber bundle and the field lens interfaces are detachable to facilitate, on one side, the attaching of the fiber input slit with the first lens of the collimator and, on the other side, the dewar assembly with the field lens in a clean and controlled environment. A focus tool consisting of a micrometer screw tool was moved from the field lens interface to the bundle interface. This new position provides a four times higher resolution than the former MUSE design.

A shutter device as depicted in Figure 4 is adapted within the gap between the first and the second lens of the collimator subassembly. The unit also supports up to three optical filter glasses to avoid second order contamination emerging from different VPHG orders of diffraction. This device is a heritage from the PMAS instrument[6,7] and consists of a metallic cylinder with a rear and front aperture that are big enough to avoid any vignetting of the input light when opened. The internal cylinder is then electronically rotated by means of a solenoid such that the input light is blocked and does not reach the detector. Moreover, two holes drilled near the field lens can be used to pump nitrogen gas into the spectrograph in order to prevent condensation of humitidy on the field lens. Most of the mechanical surfaces are black anodized, and long mechanical pieces are also threaded to avoid grazing incidence, thus preventing stray light.

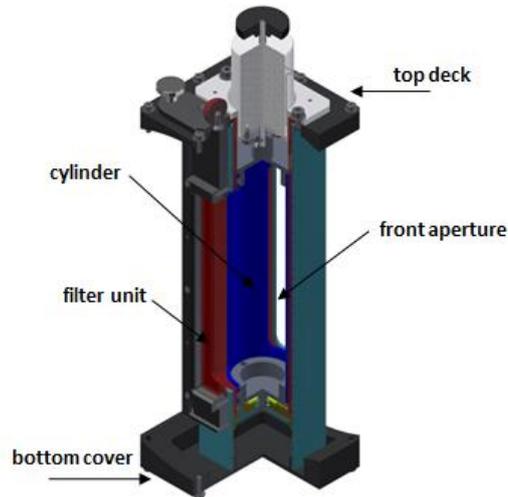

Figure 4. Shutter device and filter unit adapted from PMAS to MRS. The purpose is to block light coming on the CCD detector during readout process, and to filter out parts of the input light spectrum that are responsible of second order contamination at the image plane.

## 2.3 Detector

The detector at the image plane is housed by a custom-design aluminum case designed and manufactured by AIP. Inside, there is a 4K × 4K, 15 µm pixel back-illuminated CCD detector (e2v CCD231). Since the spectrograph has no movable parts, the detector features a graded index anti-reflection coating to enhance the quantum efficiency according to the fixed wavelength that impinges on each part of the CCD. Like in MUSE, the field lens of the spectrograph becomes the Dewar window. As a difference to the former, the high-vacuum cryogenic system is now being realized with an IGC Polycold Cryotiger cooler instead of a closed-cycle liquid nitrogen system.
.

## 3. PERFORMANCE AND TEST RESULTS

The Potsdam MRS optical performance has been tested in terms of magnification and image quality within the operational range of temperatures 15 – 25 °C. Tests were performed by Winlight Systems with tools similar to the ones that were previously used for MUSE.

### 3.1 Testing tools

Four different laser diodes (0.465 µm, 0.633 µm, 0.780 µm and 0.900 µm wavelengths) besides a mercury lamp (for the peak at 0.405 µm) were used for the evaluation. No test has been performed below 0.4 µm using the 0.365 µm mercury peak because the low efficiency of testing tools and VPHG at this ultra-violet wavelength.

To create an input object simulating a fiber with an appropriate f-number, an autocollimator coupled to a microscope objective was used, Figure 5. This device utilizes objective lenses to collimate light coming from the output of a fiber of around 0.12 NA and a second objective to focus light into the referential object plane of the spectrograph. Furthermore, the autocollimator incorporates its own camera assisting the positioning of the testing beam at the reference plane. A three axis translation stage accurately moves the device along the whole input field view with a precision of 5 µm in the vertical (X axis) and lateral axis (Y axis), and less than 25 µm in the Z axis. The magnification parameter for the autocollimator is calculated to be 0.5 and the f-number around 2 which covers the nominal input f-number of the spectrograph. At the image plane, there is an imager device, Figure 5, composed by a microscope objective coupled to a camera detector capturing conjugated image spots. Zemax simulations revealed that the highest angle at the focus plane formed by the light coming from any position of the input field at any wavelength is 30.9 degrees regarding the Z axis. Therefore, for the proper operation of the imager device, the collection objective must have a NA greater than sin (30.9°) = 0.513 in order to not vignette the beam for any of the spots imaged. Furthermore, it must be color corrected.

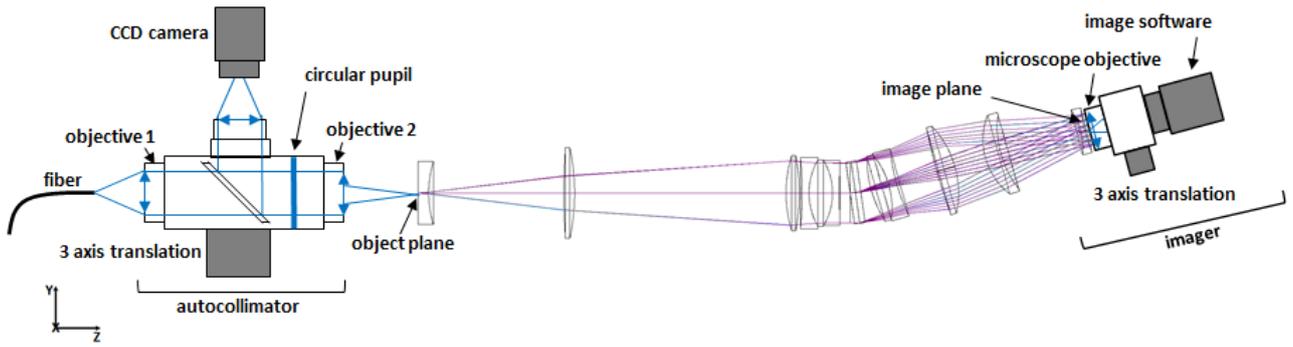

Figure 5. Optical setup utilized to test the magnification and image quality of the optical system. It is composed by an autocollimator at the input and an imager at the end of the spectrograph (not to scale).

### 3.2 Calibration of positions

Before performing the tests, the setting tools must be calibrated and the best image plane determined to get the most accurate results. For this purpose, a helium-neon laser was placed at the center of the input field of view and the image plane was centered in X and Y axis. Thereafter, the position of the laser was shifted to the borders of the field of view to check the orientation of the X axis with regard to the optical axis. The orientation of the Y axis was also tested using several wavelengths from a calibration spectral lamp located at different input positions. Finally, by means of shimming corrections the best focus plane was found.

The global position of the autocollimator and the imager devices must be also calibrated with respect to the input field of view and the image plane, respectively. A mechanical piece fixed to the slit interface defines the input aperture and two reticles at the top and bottom part indicate the relative position inside the object plane. At the image side, a piece mounted on the field lens flange with a square hole in the middle and four reticles defines and calibrates the image plane. At the borders of the camera flange, three reference pins determine the center of the reference image plane and the distance between the reticle and interface plane defines the longitudinal position of the image reference plane, Figure 6.

These calibration tools are mounted on the spectrograph and nominally set with respect to the object and image planes. Afterwards, the autocollimator points successively the referenced marks and defines their own referential system so that the planes defined by this coordinate system are the object referential plane and the image referential plane, respectively.

### 3.3 Magnification

The ratio between the input slit and the image side length defines the magnification parameter for the system. For an input slit of 118 mm and a CCD chip size of 61.44 mm, the magnification is calculated to be 0.52. This parameter only refers to the spatial axis (X axis) and must be lower or equal to the nominal value to ensure that the signals of all fibers are imaged on the detector. Magnification, G, is determined as:

$$G_x = \frac{d'}{d} = \frac{\sqrt{(X'1-X'0)^2 + (Y'1-Y'0)^2}}{\sqrt{(X1-X0)^2 + (Y1-Y0)^2}} \quad (1)$$

where X0, Y0 are the coordinates of the object created at a border of the input field, X'0, Y'0 are the referenced coordinates of the image created by this object at one wavelength and X1, Y1 are the coordinates of an object created at the opposite border of the field of view with X'1, Y'1 its image coordinates at the same wavelength.

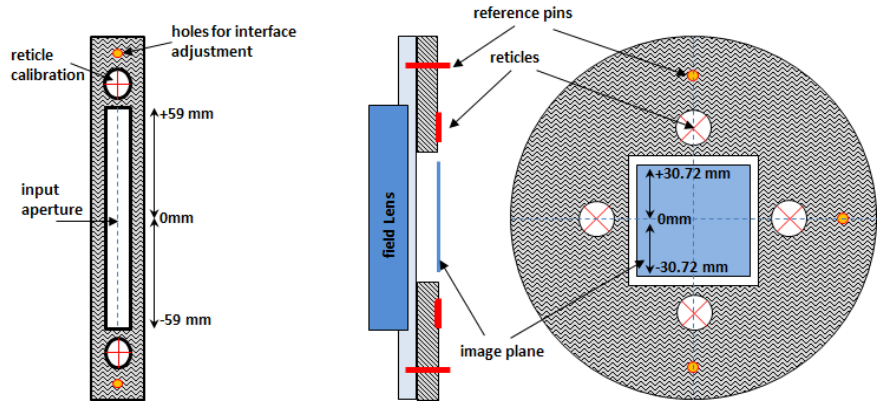

Figure 6. Mechanical device fixed at the input of the spectrograph defining the input field of view (left) and mechanical device defining the image field (right).

The magnification values at 20°C temperature for several wavelengths are shown in Table 1. According to (1), all results satisfy the requirement of G ≤ 0.52, meaning that all fibers inside the input field-of-view are imaged onto the detector.

Table 1. Magnification values for several wavelenghts at 20°C.

| Wavelength | (X0; Y0) (mm) | (X'0; Y'0) (mm) | (X1; Y1) (mm) | (X'1; Y'1) (mm) | $G_x$ |
|---|---|---|---|---|---|
| 0.405 µm | (59; 0) | (30.623; -25.349) | (-59; 0) | (-30.595; -25.381) | 0.518 |
| 0.465 µm | (59; 0) | (30.578; -17.995) | (-59; 0) | (-30.528; -18.019) | 0.517 |
| 0.633 µm | (59; 0) | (30.673; 0.439) | (-59; 0) | (-30.637; 0.475) | 0.519 |
| 0.780 µm | (59; 0) | (30.713; 16.619) | (-59; 0) | (-30.653; 16.643) | 0.520 |
| 0.900 µm | (59; 0) | (30.694; 29.425) | (-59; 0) | (-30.623; 29.438) | 0.519 |

In order to test the magnification behavior at different temperatures, the spectrograph was placed in a room with air conditioning, and the magnification tests were performed again. Table 2 and Table 3 show magnification values at 0.405 µm, 0.633 µm and 0.900 µm wavelengths at 15 °C and 25 °C, respectively. One can see that the magnification values at 15 °C and 25 °C are almost identical to the ones at 20 °C. Therefore, the athermalization of the system in terms of magnification has been proven.

Table 2. Magnification values measured at 15°C.

| Wavelength | (X0; Y0) (mm) | (X'0; Y'0) (mm) | (X1; Y1) (mm) | (X'1; Y'1) (mm) | $G_x$ |
|---|---|---|---|---|---|
| 0.405 µm | (59; 0) | (30.514; -25.364) | (-59; 0) | (-30.538; -25.271) | 0.517 |
| 0.633 µm | (59; 0) | (30.672; 0.421) | (-59; 0) | (-30.615; 0.452) | 0.519 |
| 0.900 µm | (59; 0) | (30.615; 29.547) | (-59; 0) | (-30.59; 29.55) | 0.518 |

Table 3. Magnification values measured at 25°C.

| Wavelength | (X0; Y0) (mm) | (X'0; Y'0) (mm) | (X1; Y1) (mm) | (X'1; Y'1) (mm) | $G_x$ |
|---|---|---|---|---|---|
| 0.405 µm | (59; 0) | (30.562; -25.393) | (-59; 0) | (-30.545; -25.309) | 0.517 |
| 0.633 µm | (59; 0) | (30.670; 0.410) | (-59; 0) | (-30.625; 0.470) | 0.519 |
| 0.900 µm | (59; 0) | (30.601; 29.512) | (-59; 0) | (-30.605; 29.501) | 0.518 |

## 3.4 Image quality

The diffraction ensquared energy (EE) criterion was selected for the assessment of image quality. It is expected that energy enclosed in 30 µm × 30 µm square area (2 × 2 pixels of the 4K × 4K CCD detector) to be higher than 80% of the first order energy diffracted by the VPHG.

For the evaluation, a simulated object is moved by the three axis translation stage along predetermined positions within the input field of view. At the image plane, the imager device records the point spread function (PSF) for several wavelengths within the nominal spectral range from any object position. Each spot and its parameters were digitally stored and then a computer software calculated the EE diffracted inside a specified area. The process was repeated for the different input object positions.

The size of the spots at the image plane is too small to be properly evaluated without any enlargement. A rough estimation shows that the image of the fiber core utilized as object for the autocollimator at the input of the spectrograph is around 10 µm. Since the magnification is 0.52, the image spots at the detector side are expected to be around 5 µm diameter which is smaller than the image quality requirement. An appropriate evaluation of the PSF and EE can be realized by using a 20× magnification objective for the imager device to enlarge the spots and a 6 µm x 6 µm pixel camera to provide better image resolution. By means of this transformation, a 30 µm × 30 µm area is now represented in 100×100 pixels at the imager detector (30 µm × 20 magnification / (6 µm / pixel)) = 100 pixels) and the geometrical distribution of energy can be better estimated. It must be noticed that owing to the broad range of wavelengths covered at the image plane, two different cameras were utilized to record the imaged spots with an acceptable signal-to-noise quality. In particular, a camera model Scout sca640-70gm manufactured by Basler (Ahrensburg, Germany) records spots between 0.465 µm and 0.9 µm wavelengths whilst a camera model ST-8300M manufactured by Diffraction Limited / SBIG (Otawa, Canada) images spots only at 0.405 µm wavelength.

The EE values measured in a nominal area of 30 × 30 µm$^2$ are shown in Table 4. Five wavelengths and five object positions were chosen for the test. All measured values at 20 °C are above EE ≥ 80 % and most of the best EE values are achieved for objects created at 0 mm input position. There are unexpected discrepancies between EE values at 0.465 µm and 0.405 µm, likely due to the use of two different cameras and the sensitivity of the EE algorithm to detector noise.

Table 4. Ensquared energy values measured at 20 °C in 30 × 30 µm$^2$ area for different wavelengths and objects.

| Wavelenght/Field | -59 mm | -40 mm | 0 mm | 40 mm | 59 mm |
|---|---|---|---|---|---|
| 0.900 µm | 86.7 % | 87.3 % | 85.5 % | 90.5 % | 92.2 % |
| 0.780 µm | 88.5 % | 89.9 % | 93.0 % | 93.7 % | 90.1 % |
| 0.633 µm | 89.7 % | 90.2 % | 92.4 % | 88.1 % | 88.0 % |
| 0.465 µm | 82.5 % | 87.3 % | 89.0 % | 83.7 % | 81.6 % |
| 0.405 µm | 87.5 % | -- | 93.2 % | -- | 87.4 % |

The Point Spread Function (PSF) images recorded for the calculation of EE in Table 4 are shown in Figure 7. The upper part of the figure refers to the PSF captured with the infrared camera while the blue enhanced camera recorded the three bottom spots. In both cases, software calculates the energy enclosed inside the light blue square which delimits the nominal 30 × 30 µm$^2$ area.

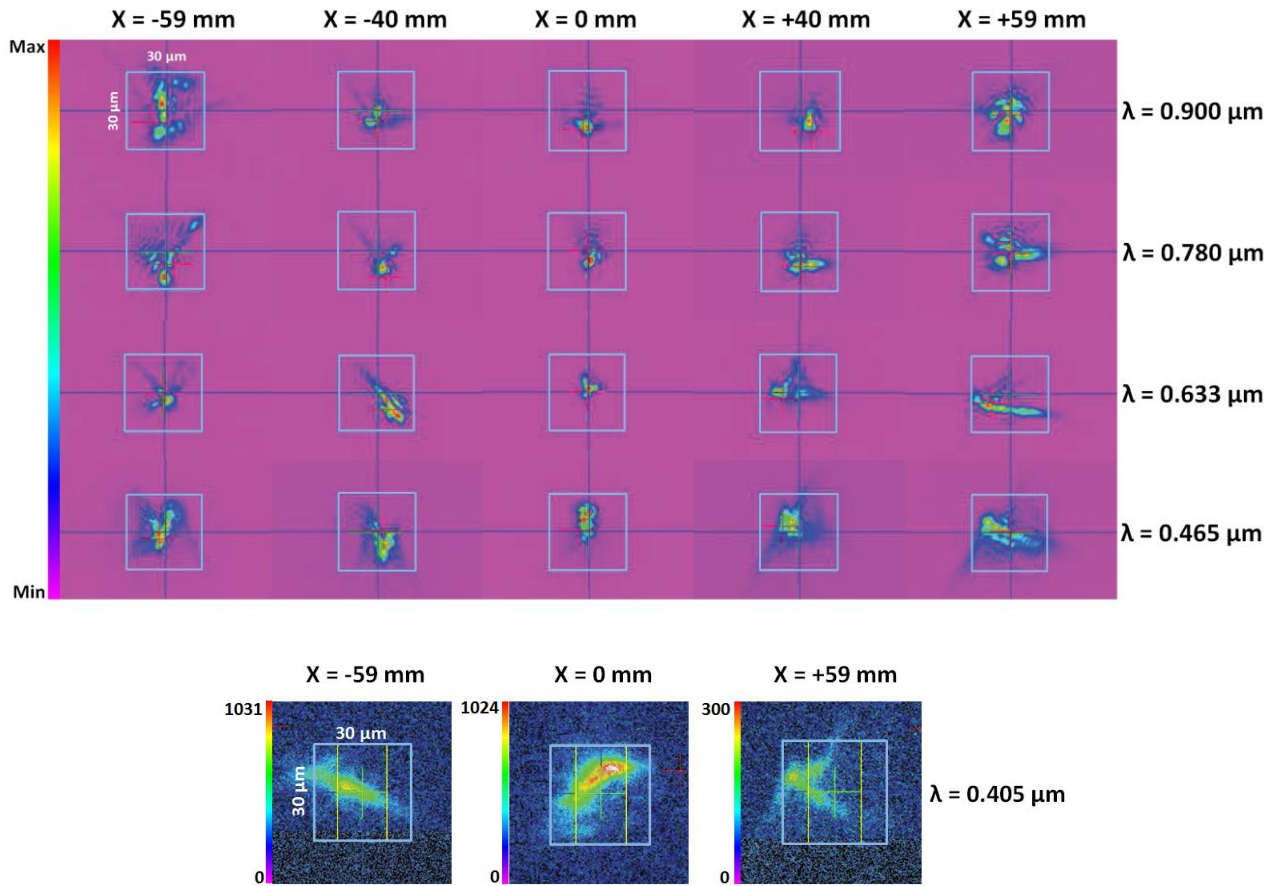

Figure 7. PSF recorded with red sensitive camera at five object positions for four wavelengths (top). PSF captured with blue sensitive camera for 0.405μm wavelength (bottom).

The PSF was also recorded at 15 °C and 25 °C for three object positions (-59 mm, 0 mm and +59 mm) and three wavelengths (0.405 µm, 0.633 µm and 0.9 µm). Table 5 shows EE values at 25 °C and Table 6 values at 15 °C. One can see with regard to temperature that the image quality is maintained for all tested wavelengths and object points.

Table 5. EE values calculated at 25ºC.

| Wavelenght/Field | -59 mm | 0 mm | 59 mm |
|---|---|---|---|
| 0.405 µm | 81.5 % | 89 % | 85.8 % |
| 0.633 µm | 88.8 % | 90.4 % | 84.9 % |
| 0.900 µm | 83.1 % | 82.7 % | 93.3 % |

Table 6. EE values calculated at 15 °C.

| Wavelenght/Field | -59 mm | 0 mm | 59 mm |
|---|---|---|---|
| 0.405 µm | 87.6 % | 86.7 % | 85.0 % |
| 0.633 µm | 88.2 % | 94.0 % | 87.9 % |
| 0.900 µm | 85.8 % | 83.1 % | 92.2 % |

Figure 8, Figure 9 and Figure 10 show a comparison regarding PSF images at three object positions at 25 °C and 15 °C temperatures for 0.405 µm, 0.633 µm and 0.9 µm wavelengths, respectively.

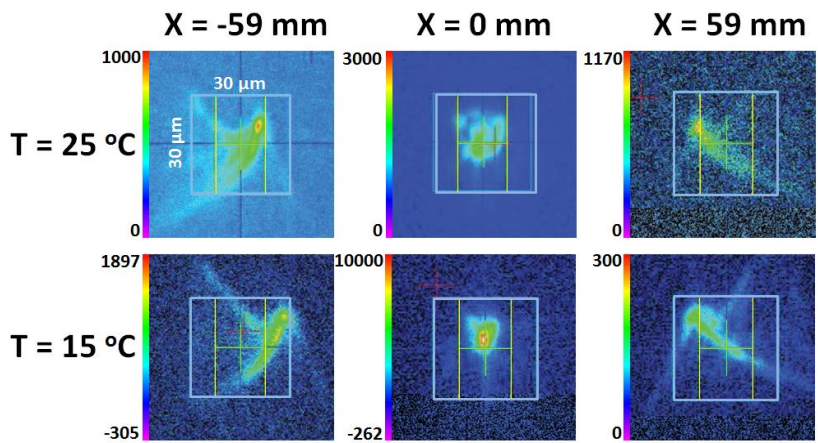

Figure 8. PSF captured with the blue sensitive camera for three object positions and 0.405 µm wavelength at 25 °C and 15 °C temperatures.

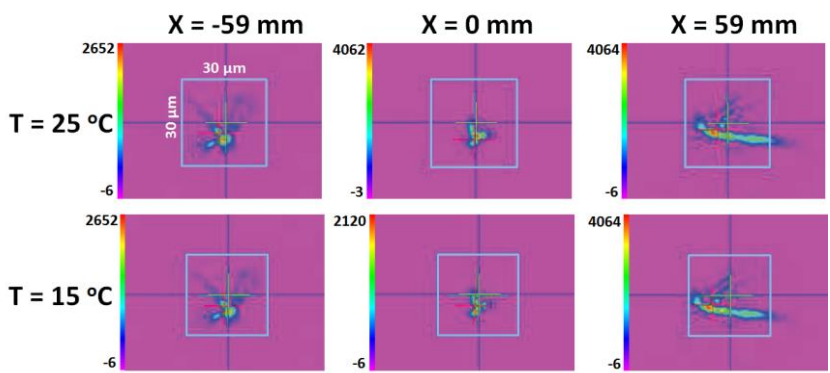

Figure 9. PSF captured with the red sensitive camera for three object positions and 0.633 μm wavelength at 25 °C and 15 °C temperatures.

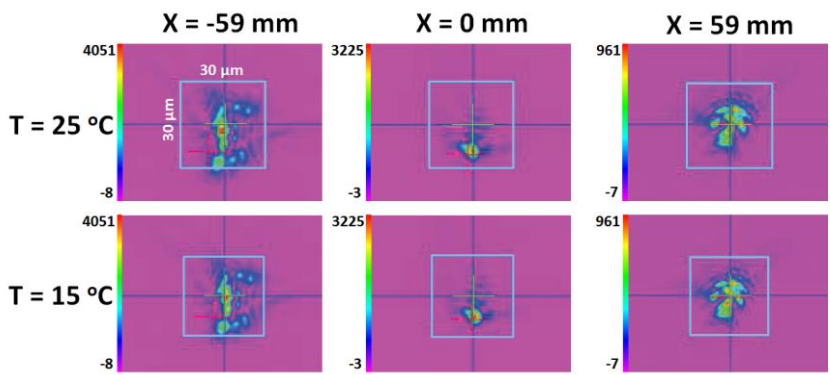

Figure 10. PSF captured with the red sensitive camera for three object positions and 0.900 µm wavelength at 25 °C and 15 °C temperatures.

## 4. PRELIMINARY RESULTS FROM FIRST-LIGHT

The fiber-slit unit and the camera have been recently assembled with the spectrograph in a provisional setup to obtain first laboratory light on May 26, 2016. This configuration is preliminary in the sense that the detector was not yet mounted in the final detector head, resulting in an offset of the detector from the nominal focal plane of about 1.45 mm. As a result, in a series of focus exposures the minimum of FWHM is not reached. Therefore the system could not yet be tested at best focus. Also, only 396 out of 400 fibers are imaged on the detector. Nevertheless, one can already obtain a qualitative impression of the performance. The fibers used for this experiment have a core size of 114 μm, resulting in a projected image diameter of 59 μm on the detector. In Figure 11 the intensity profile along the cross-dispersion axis of a group of 20 fibers near the optical axis is shown. The wavelength corresponding to this position is near 0.640 μm. The FWHM of the spectra in the cross-dispersion direction is qualitatively in agreement with the expectation of 4 pixels. This is also seen in Figure 12 which shows a spot at 0.640 μm coming from one of the 20 fibers illuminated with a neon lamp.

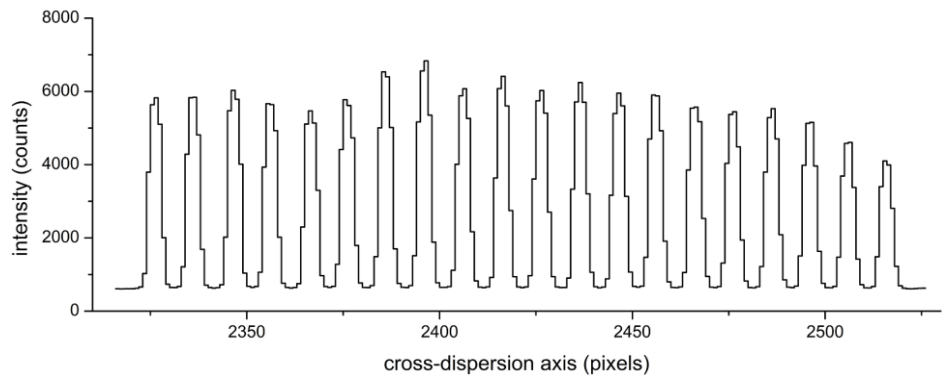

Figure 11. Intensity profile along the cross-dispersion axis for a group of 20 centered fibers at 0.640 μm wavelength.

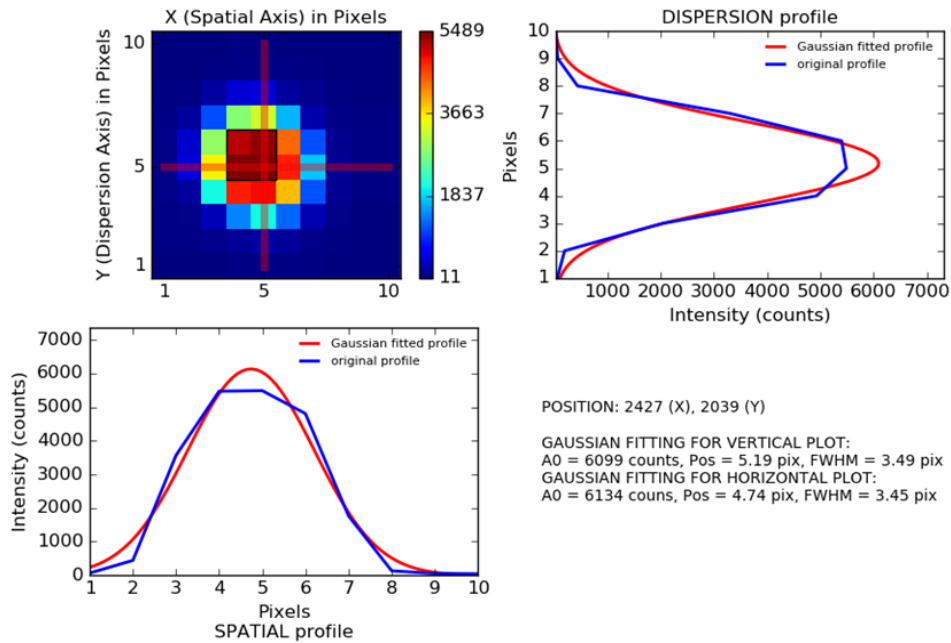

Figure 12. Intensity profile along the dispersion and spatial axis for a neon spot at 0.640 μm created for a more or less centered fiber.

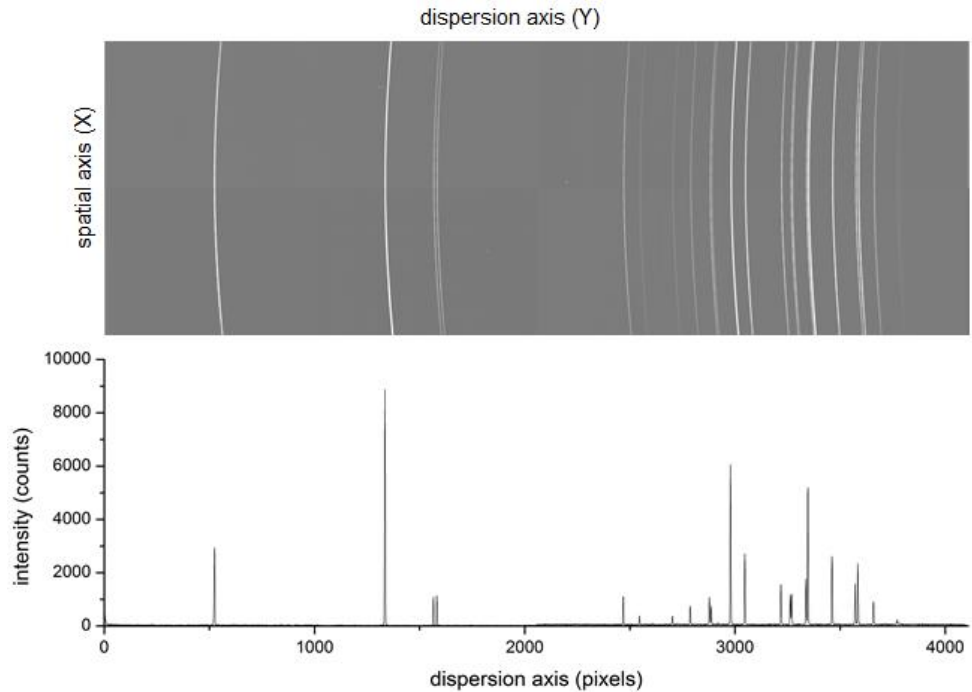

Figure 13. P3D output after ray tracing detection (top) and spectrum from a mercury lamp (bottom) recorded as end-to-end measurement using the MRS spectrograph. The left part and the right part of the spectrum correspond to blue and red wavelengths, respectively.

The data reduction for this instrument will be performed using the p3d software[8] which is an open source general purpose package for fiber-coupled IFU spectroscopy (http://p3d.sourceforge.net/). In the absence of the exact geometry of the final hardware configuration (detector head) the p3d configuration file for the MRS spectrograph has not yet been created. However, the basic functionality, e.g. BIAS subtraction and extraction of spectra, is already being utilized for initial inspection. An example of an extracted mercury arc spectrum is shown in the Figure 13.

## 5. CONCLUSIONS

The opto-mechanical design of the Potsdam Multiplex-Raman-Spectrograph has been presented and test results of the optical performance have been reported. The design of this optical system is based on the successful MUSE spectrograph, however fed with optical fibers instead of image slicers, featuring an extended blue wavelength coverage. Despite of being constructed for the purpose of laboratory Raman spectroscopy, it can be also used in astronomy for IFS and multi-object-spectroscopy. As part of the mechanical modifications, the spectrograph incorporates an integrated shutter device. First order diffraction measurement tests have been carried out with the VPHG that indicate good diffraction efficiency, especially at blue wavelengths. The optical performance of the whole spectrograph in terms of magnification and image quality has been evaluated and its calibration and test procedures have been explained. Results show that the magnification was maintained at values close to nominal for all input objects and wavelengths tested at temperatures of 25 °C, 20 °C and 15 °C, while image quality, evaluated in terms of ensquared energy within 30 µm × 30 µm, shows an EE higher than 80 %. The athermalization of the system in the range of temperatures from 15 °C to 25 °C has been demonstrated. Provisional first light end-to-end measurements show that qualitatively the system works as designed.

## ACKNOWLEDGEMENTS

innoFSPEC Potsdam is supported by the BMBF program "Unternehmen Region" under grant no. 03Z2AN11. BM and ES acknowledge support from the BMBF VIP program under grant no. 03V0843. The authors are indebted to Joss Bland-Hawthorn and Roland Bacon for having initially proposed for the FIREBALL project the now realized fiber-coupled version of the MUSE spectrograph design.